\documentclass[a4paper,11pt]{article}

\usepackage{amsmath,amssymb,color,graphics,epsfig,cite}
\usepackage{titlesec}
\usepackage[titletoc,title]{appendix}

\usepackage{amsfonts}
\newcommand{\hoch}[1]{$\, ^{#1}$}
\newcommand{\dd}{\mathrm{d}}
\newcommand{\be}{\begin{equation}}
	\newcommand{\ee}{\end{equation}}
\newcommand{\bea}{\setlength\arraycolsep{2pt} \begin{eqnarray}}
	\newcommand{\eea}{\end{eqnarray}}

\def\0{{\sst{(0)}}}
\def\1{{\sst{(1)}}}
\def\2{{\sst{(2)}}}
\def\3{{\sst{(3)}}}
\def\4{{\sst{(4)}}}
\def\5{{\sst{(5)}}}
\def\6{{\sst{(6)}}}
\def\7{{\sst{(7)}}}
\def\8{{\sst{(8)}}}
\def\sst#1{{\scriptscriptstyle #1}}

\begin{document}
	\begin{center}
	{\Large {\bf Matter Maps to Geometry in Gravitational Collapse}} 
	
	\vspace{20pt}	
H. Khodabakhshi\hoch{1}, Zhi-Chao Li\hoch{1}, H. L\"u\hoch{1,2}, F. Shojai\hoch{3}
	
\vspace{20pt}

{\it \hoch{1}Center for Joint Quantum Studies and Department of Physics,\\
	School of Science, Tianjin University, Tianjin 300350, China }
\medskip

{\it \hoch{2}The International Joint Institute of Tianjin University, Fuzhou,\\ Tianjin University, Tianjin 300350, China}

\medskip
{\it \hoch{3}Department of Physics, University of Tehran, P.O. Box 14395-547, Tehran, Iran}

		\vspace{40pt}
	
\underline{ABSTRACT}
\end{center}

	We establish an exact bidirectional map between the effective homogeneous interior density of a collapsing star and the exterior metric function in generalized Oppenheimer--Snyder collapse. The Darmois--Israel junction conditions reduce this relation to a purely algebraic form, providing a direct way to reconstruct candidate static geometries and their surface dynamics without solving the corresponding differential field equations separately. Correction powers diagnose models: integer exponents signal ultraviolet completions, while fractional powers identify phenomenological ones. Our framework not only simplifies the construction of regular black holes but also provides a universal benchmark for testing singularity resolution and cosmic censorship in quantum gravity phenomenology.

\vfill {\footnotesize  h\_khodabakhshi@tju.edu.cn \ \ \ 
lizc@tju.edu.cn \ \ \
mrhonglu@gmail.com\ \ \ fshojai@ut.ac.ir}

\thispagestyle{empty}
\pagebreak

\tableofcontents
\addtocontents{toc}{\protect\setcounter{tocdepth}{2}}

\section{Introduction}

Gravitational collapse need not end in a singularity. In the context of Oppenheimer--Snyder collapse~\cite{Oppenheimer:1939} and its generalization to arbitrary static exteriors~\cite{Khodabakhshi:2025dya}, an exact bidirectional mapping between interior matter content and exterior geometry provides a direct framework for generating black hole solutions and determining their collapse dynamics. By inserting physically motivated corrections into the total density $\rho$ in the Friedmann equation, one can reconstruct the corresponding geometry and surface evolution. 

Crucially, we emphasize from the outset that the restriction to the $g_{tt}g_{rr} = -1$ exterior class is not an \textit{a priori} assumption, but rather the strict mathematical consequence of demanding a smooth, force-free Darmois--Israel junction for a homogeneous interior. We prove this explicitly in Sec. 2. 

While our prior work~\cite{Khodabakhshi:2025dya,Li:2026ub,Li:2026OS} established the generalized Oppenheimer--Snyder (GOS) framework, classified collapse outcomes into singular/bouncing/soft-landing scenarios, and conjectured an upper bound on horizon formation time (saturated by Schwarzschild), the present article introduces three new elements: 
(i) an exact bidirectional algebraic mapping that generates solutions without solving differential equations, (ii) a diagnostic criterion based on the analytic structure of density corrections to distinguish systematic UV-motivated expansions from phenomenological models, and (iii) an explicit construction demonstrating that loop quantum gravity (LQG)-corrected black hole geometries admit exact classical realizations in nonlinear electrodynamics with identical collapse dynamics.

The force-free Israel junction conditions~\cite{Israel1966} restrict the exterior to the special class of static, spherically symmetric exterior metrics with $g_{tt}g_{rr} = -1$~\cite{Li:2026ub}, leading to the star's surface evolution equation
\begin{equation}
	\dot R^2 = 1 - f(R),
	\label{eq:surface_law}
\end{equation}
where the dot denotes differentiation with respect to the proper time $T$, the collapsing branch corresponds to $\dot R < 0$, and real motion requires $1 - f(R) \geq 0$. The interior homogeneous effective fluid satisfies an effective Friedmann equation $H^2 = 8\pi\rho/3$ with $H = \dot R/R$, where $\rho(R)$ is the star's total effective density including standard matter together with all corrections from quantum effects, higher-derivative terms, or other physics. Combining these yields
\begin{equation}
	\rho(R) = \frac{3}{8\pi R^2} \bigl[ 1 - f(R) \bigr],
	\label{eq:main_mapping}
\end{equation}
where $\rho(R)$ decomposes into standard contributions $\rho_{\text{s}}$ and effective corrections $\rho_{\text{eff}}$ from quantum or modified gravity effects. The standard contribution $\rho_{\text{s}}$ consists of a neutral dust term, a radiation-like effective term associated with the exterior charge via the mapping, and a cosmological constant term. The surface evolution equation~\eqref{eq:surface_law} and the algebraic mapping~\eqref{eq:main_mapping} follow directly from the Darmois--Israel junction conditions; a complete, coordinate-independent derivation is provided in Sec. 2. The relation is fully bidirectional: prescribing $\rho(R)$ reconstructs the exterior geometry $f(R)$, while any static solution $f(R)$ determines the effective density sourcing the collapse. Equation~\eqref{eq:main_mapping} thereby reduces the differential matter--geometry relation of Einstein's field equations to a much simpler purely algebraic correspondence. Unlike Hamiltonian reconstruction~\cite{ZhangCao2025}, our method derives directly from the junction conditions.

The algebraic map is not tied to a particular microscopic matter model. In the Reissner--Nordstr\"om (RN) case, it describes a neutral homogeneous interior whose boundary carries the charge sourcing the exterior Maxwell field; in other examples it can be used as a reconstruction tool for effective density profiles or as a diagnostic for candidate exterior geometries. By inserting physically motivated corrections to the effective density $\rho_{\text{eff}}$---such as those from loop quantum cosmology (LQC)~\cite{Ashtekar:2006uz} or quasi-topological gravity~\cite{Bueno:2025}---one immediately obtains the corresponding black hole geometry. The structure of $\rho_{\rm eff}(R)$ also helps diagnose whether a geometry is associated with a systematic UV-motivated expansion or is more naturally phenomenological. When corrections admit a Taylor expansion in integer powers of $\rho_{\rm m}$---as in systematic UV-motivated perturbative expansions such as LQG or quasi-topological gravity---the resulting metric function $f(R)$ contains only terms of the form $R^{-(3n-2)}$ with $n \in \mathbb{N}$. In contrast, phenomenological models like the Bardeen black hole~\cite{Bardeen1968} typically involve non-analytic or fractional powers of $R$, indicating an effective rather than fundamental origin. Generic quantum corrections may instead generate time-dependent exteriors, inhomogeneous interiors, nonlocal effective stresses, or surface layers, and such cases lie outside the scope of the present work.

In Ref.~\cite{Li:2026OS}, we showed that the shape of $f(R)$ determines whether collapse ends in a singularity, bounces at finite radius, or approaches the center only asymptotically---leading to the three scenarios: \textit{singular}, \textit{bouncing}, and \textit{soft-landing}. Here, we use this classification as a diagnostic for the mapping \eqref{eq:main_mapping}. In Appendix A, we construct an exact nonlinear electrodynamics (NLED) counterpart of an LQG-corrected black hole~\cite{Kelly:2020lec, Ali:2022lec, Lewandowski:2023lec,AyonBeato:1998}. With matched parameters, both yield identical $R(T)$ and apparent-horizon evolutions. Similarly, the Hayward metric~\cite{Hayward:2005gi} exhibits soft-landing collapse with a de Sitter core, while its large-$R$ expansion shares the same leading correction as the LQG-inspired metric, illustrating that identical leading asymptotic corrections need not imply identical global collapse dynamics. We work in units $G = c = 1$ and metric signature $(-,+,+,+)$.

Finally, we emphasize the broader implications of this framework. While the force-free GOS model applies to a specific subset of collapse scenarios, the exact algebraic mapping derived herein serves as a universal, non-perturbative benchmark for quantum gravity phenomenology. Any complete theory of quantum collapse must either reduce to this exact force-free sector in the homogeneous limit or explicitly account for the thin shells and inhomogeneities it departs from. Thus, our results provide a rigorous testing ground for singularity resolution and cosmic censorship across a wide class of modified gravity and quantum geometry models, extending the significance of this work well beyond the specific ansatz.

\section{Explicit derivation of the junction conditions}
\label{sec:junction}

\noindent The Darmois--Israel junction conditions are geometric statements that are independent of the coordinate system used to evaluate them. We present the complete derivation here, following the matching procedure established in Ref.~\cite{Khodabakhshi:2025dya} and refined in Ref.~\cite{Li:2026ub}, to demonstrate rigorously how Eqs.~\eqref{eq:surface_law} and \eqref{eq:main_mapping} follow from the continuity of the induced metric and extrinsic curvature.

Consider a spatially flat FLRW interior matched to a general static, spherically symmetric exterior across a timelike hypersurface $\Sigma$ representing the star surface. In standard coordinates, the exterior metric is
\begin{equation}
	ds^2_{\rm out} = -h(r)\, dt^2 + \frac{dr^2}{f(r)} + r^2 d\Omega^2,
	\label{eq:static_out_app}
\end{equation}
while the interior FLRW metric in Painlev\'e--Gullstrand form is
\begin{equation}
	ds^2_{\rm in} = -d\tau^2 + \bigl(dr - r H(\tau) d\tau\bigr)^2 + r^2 d\Omega^2,
	\label{eq:PG_in_app}
\end{equation}
where $\tau$ is the proper time of comoving observers and $H(\tau) = \dot{a}/a = \dot{r}/r$. The surface $\Sigma$ is defined by $r = R(\tau)$ and is parametrized by $\tau$.

The first junction condition requires continuity of the induced metric $h_{ab}$ across $\Sigma$. Restricting the exterior line element to $\Sigma$ gives $ds^2_\Sigma = [-h(R)\dot{t}^2 + f(R)^{-1}\dot{R}^2] d\tau^2 + R^2 d\Omega^2$, where dots denote $d/d\tau$. Continuity with the interior induced metric $ds^2_\Sigma = -d\tau^2 + R^2 d\Omega^2$ yields the temporal relation
\begin{equation}
	-h(R)\dot{t}^2 + \frac{\dot{R}^2}{f(R)} = -1.
	\label{eq:1st_jc_temporal}
\end{equation}
This equation relates the exterior coordinate time $t$ to the interior proper time $\tau$ but remains undetermined until the second junction condition is imposed.

The second junction condition requires continuity of the extrinsic curvature $K_{ab} = e^\mu_a e^\nu_b \nabla_\mu n_\nu$ across $\Sigma$. In the absence of a thin shell ($[K_{ab}]=0$), the nontrivial mixed-index components $K^a_b = h^{ac}K_{cb}$ evaluated on $\Sigma$ are
\begin{align}
	K^{\theta}_{\theta} = K^{\phi}_{\phi} &= \frac{1}{R} \sqrt{f(R) + \dot{R}^2}, \\
	K^{\tau}_{\tau} &= \frac{1}{2\sqrt{f(R) + \dot{R}^2}} \left[ 2\ddot{R} + \bigl(f(R) + \dot{R}^2\bigr) \frac{h'(R)}{h(R)}-\dot{R}^2 \frac{f'(R)}{f(R)} \right],
\end{align}
where primes denote $d/dr$. For the spatially flat FLRW interior with a comoving boundary, the corresponding components are
\begin{equation}
	K^{\theta}_{\theta} = K^{\phi}_{\phi} = \frac{1}{R}, \qquad K^{\tau}_{\tau} = 0.
\end{equation}

Matching the angular components $[K^{\theta}_{\theta}]_\Sigma = 0$ immediately yields Eq.~\eqref{eq:surface_law}. Differentiating Eq.~\eqref{eq:surface_law} with respect to $\tau$ gives $\ddot{R} = -\frac{1}{2} f'(R)$. Substituting $\dot{R}^2 = 1-f(R)$ and $\ddot{R} = -f'(R)/2$ into the temporal matching condition $[K^{\tau}_{\tau}]_\Sigma = 0$ leads directly to
\begin{equation}
	\frac{h'(r)}{h(r)} = \frac{f'(r)}{f(r)}.
	\label{eq:h_f_relation_app}
\end{equation}
Integration yields $h(r) = C f(r)$, where $C$ is a constant. For an asymptotically flat exterior, the standard normalization $h(\infty)=f(\infty)=1$ fixes $C=1$, and thus $h(r) = f(r)$. This demonstrates that the special static condition $g_{tt}g_{rr}=-1$ is a strict consistency requirement for force-free matching, not an ansatz.

With $h=f$ established, the exterior metric can be transformed to infalling Painlev\'e--Gullstrand (PG) coordinates via $dT = dt + \frac{\sqrt{1-f}}{f} dr$. The line element becomes
\begin{equation}
	ds^2_{\rm out} = -dT^2 + \bigl(dr + \sqrt{1-f(r)}\, dT\bigr)^2 + r^2 d\Omega^2.
\end{equation}
The identification $T \equiv \tau$ follows directly from the first junction condition~\eqref{eq:1st_jc_temporal}. Substituting $h=f$ and $\dot{R}^2 = 1-f(R)$ into~\eqref{eq:1st_jc_temporal} gives $\dot{t} = 1/f(R)$. Along the surface, $dT/d\tau = \dot{t} + \frac{\sqrt{1-f}}{f} \dot{R}$. Using the collapsing branch $\dot{R} = -\sqrt{1-f}$, we obtain $dT/d\tau = \frac{1}{f} - \frac{1-f}{f} = 1$. Hence $T \equiv \tau$ is a direct consequence of the matching conditions and the geometric definition of PG time, not an independent assumption. This identification holds for any homogeneous interior whose density satisfies the junction conditions.

Finally, the interior homogeneous fluid satisfies the Friedmann equation $H^2 = 8\pi\rho/3$ with $H = \dot{R}/R$. Substitution of Eq.~\eqref{eq:surface_law} into this equation yields the algebraic mapping defined in Eq.~\eqref{eq:main_mapping}. The radial pressure continuity $[p_r]_\Sigma = 0$ follows identically from the interior energy conservation law $\dot{\rho} + 3H(\rho + p_{\rm in}) = 0$ together with Eq.~\eqref{eq:main_mapping}, ensuring a smooth junction without a thin shell.

\section{Mapping collapse dynamics to static geometries} 

In the GOS framework, force-free matching requires continuity of radial stress across the star's surface $\Sigma$. Interpreting the exterior metric as sourced by an anisotropic fluid with $T^\mu{}_\nu = \mathrm{diag}(-\rho_{\rm out}, p_r, p_t, p_t)$, the exterior energy density and pressures are \cite{Khodabakhshi:2025dya}
\begin{align}
	\rho_{\rm out}(R) &= -p_{r,\rm out}(R) = \frac{1 - f(R) - R f'(R)}{8\pi R^2}, \\
	p_{t,\rm out}(R) &= \frac{1}{8\pi}\left( \frac{f'(R)}{R} + \frac{f''(R)}{2} \right).
	\label{eq:pr_rho_pt_general}
\end{align}
The interior effective pressure follows from energy conservation $\dot\rho + 3H(\rho + p_{\rm in}) = 0$, yielding
\begin{equation}
	p_{\rm in}(R) = -\frac{\dot\rho}{3H} - \rho = p_{r,\rm out}(R),
	\label{eq:p_int}
\end{equation}
where we used the mapping \eqref{eq:main_mapping} and the surface equation \eqref{eq:surface_law}. Thus, the mapping automatically ensures force-free matching: no thin shell is required, and the junction is smooth by construction. Furthermore, the identification $T \equiv \tau$ follows directly from the first junction condition and the proper-time parametrization of $\Sigma$, as derived in Sec. 2.

The effective density decomposes naturally as
\begin{equation}
	\rho(R) = \rho_{\text{s}}(R) + \rho_{\text{eff}}(R), \quad
	\rho_{\text{s}}(R) = \frac{3}{4\pi}\left(\frac{\alpha}{R^3} + \frac{\beta}{R^4} + \gamma\Lambda\right),
	\label{eq:rho_decomp}
\end{equation}
where $\rho_{\text{s}}$ contains standard contributions (neutral dust, electromagnetic field energy, and dark energy) and $\rho_{\text{eff}}$ encodes corrections from quantum or modified gravity effects. The $R^{-3}$ term corresponds to neutral dust density $\rho_{\text{m}} = 3M/(4\pi R^3)$. The $R^{-4}$ term is associated with the electromagnetic field of the RN exterior, sourced by a charge carried by the star surface $\Sigma$, not by a charge density distributed throughout the FLRW bulk. The interior region remains neutral and homogeneous. Thus the $R^{-4}$ term in the generalized OS map represents the repulsive surface-supported electromagnetic contribution entering the boundary dynamics, not a homogeneous charged fluid inside the star. An exact FLRW interior is homogeneous and isotropic, whereas a nonzero radial electromagnetic field in the bulk would generate anisotropic stresses,
	\begin{equation}
		p_{r, \rm EM}=-\rho_{\rm EM}, \qquad p_{t,\rm EM}=+\rho_{\rm EM},
	\end{equation}
	in conflict with the isotropic FLRW stress tensor. The discontinuity of the normal electromagnetic field across the boundary implies an idealized surface charge current through the Maxwell junction condition,
	\begin{equation}
		n_{\mu}\left(F_{\rm out}^{\mu\nu} - F_{\rm in}^{\mu\nu}\right)=4\pi j_{\Sigma}^{\nu}.
	\end{equation}
	Since $F_{\rm in}^{\mu\nu}=0$ in the neutral FLRW interior, while the exterior RN field is nonzero, the boundary carries a nonvanishing surface current. For a spherically symmetric surface with total charge $q$, its surface charge density is
	\begin{equation}
		\sigma_{q}=\frac{q}{4\pi R^{2}}.
	\end{equation}
	This electromagnetic surface current should not be confused with an Israel thin shell. A gravitational thin shell would require a nonzero surface stress-energy tensor,
	\begin{equation}
		S_{ab}=-\frac{1}{8\pi} \left( [K_{ab}]-h_{ab}[K]\right).
	\end{equation}
	In the present construction, the induced metric and extrinsic curvature are continuous, so
	\begin{equation}
		[K_{ab}]=0, \qquad S_{ab}=0.
	\end{equation}
	Thus, the boundary carries an idealized Maxwell surface current but no Israel surface stress-energy layer.

For $\rho_{\text{eff}} = 0$, substituting \eqref{eq:rho_decomp} into \eqref{eq:main_mapping} reproduces the RN--(A)dS metric
\begin{equation}
	f_{\text{RN}}(R) = 1 - \frac{2M}{R} + \frac{q^2}{R^2} - \frac{\Lambda}{3}R^2,
	\label{eq:f_RN}
\end{equation}
with coefficients fixed by matching:
\begin{equation}
	\alpha = M, \qquad \beta = -\frac{q^2}{2}, \qquad \gamma = \frac{1}{6}.
	\label{eq:coeff_map}
\end{equation}

Since $\beta = -q^2/2 < 0$, the $R^{-4}$ term represents a negative radiation-like contribution to the homogeneous effective interior density,
	\begin{equation}
		\rho_{\text{rad}}(R)= \frac{3\beta}{4\pi R^{4}} = -\frac{3q^{2}}{8\pi R^{4}}.
	\end{equation}
	This negative effective energy density should be contrasted with the exterior Maxwell energy density, which remains positive,
	\begin{equation}
		\rho_{\rm EM, out}(R)=\frac{q^{2}}{8\pi R^{4}}.
	\end{equation}
	The negative radiation-like term also ensures continuity of the radial stress across the star boundary, while the Maxwell surface current remains distinct from an Israel surface stress-energy layer. From the interior conservation equation,
$
		\dot{\rho}_{\rm in} + 3H \left( \rho_{\rm in}+p_{\rm in} \right)=0,
$
	and using $\dot{R}=HR$, the effective interior pressure can be written as
	\begin{equation}
		p_{\rm in}(R)= -\rho_{\rm in}(R)-\frac{R}{3} \frac{d\rho_{\rm in}(R)}{dR}.
	\end{equation}
	For the density determined by Eqs.~\eqref{eq:rho_decomp} and \eqref{eq:coeff_map},
	\begin{equation}
		\rho_{\rm in}(R)= \frac{3M}{4\pi R^{3}} - \frac{3q^{2}}{8\pi R^{4}} +\frac{\Lambda}{8\pi},
	\end{equation}
	one obtains
	\begin{equation}
		p_{\rm in}(R)= -\frac{q^{2}}{8\pi R^{4}} - \frac{\Lambda}{8\pi}.
	\end{equation}
	This is exactly equal to the exterior radial pressure,
	\begin{equation}
		p_{r,\rm out}(R)= -\frac{q^{2}}{8\pi R^{4}} -\frac{\Lambda}{8\pi}.
	\end{equation}
	Hence,
	\begin{equation}
		p_{\rm in}\big|_{\Sigma}=p_{r,\rm out}\big|_{\Sigma},
	\end{equation}
	and the radial stress is continuous across the boundary. No Israel thin shell is therefore required. The negative $R^{-4}$ term should be understood as a homogeneous effective contribution that enforces force-free gravitational matching, while the charge itself is represented by the Maxwell surface current. Setting $\beta = \gamma = 0$ recovers the Schwarzschild solution. This RN example should therefore be read as a neutral FLRW interior matched to an RN exterior whose charge is carried by the star surface. It is not a model of a homogeneous charged FLRW fluid. The distinction is important: a bulk charged fluid in Einstein--Maxwell theory would generally be inhomogeneous and anisotropic, whereas a surface charge produces the exterior RN field while preserving the neutrality and homogeneity of the FLRW bulk interior.

The mapping \eqref{eq:main_mapping} operates bidirectionally: (i) prescribing $\rho(R)$ reconstructs $f(R)$ algebraically; (ii) given $f(R)$, one extracts the effective density sourcing the collapse. Its power becomes evident when incorporating corrections.

\subsection{From density to geometry.} LQC replaces the classical density with the Ashtekar--Pawlowski--Singh (APS) form~\cite{Ashtekar:2006uz}
\begin{equation}
	\rho_{\text{APS}}(R) = \rho_{\text{m}}\left(1 - \frac{\rho_{\text{m}}}{\rho_c}\right), \qquad \rho_{\text{m}} = \frac{3M}{4\pi R^3},
	\label{eq:rho_APS}
\end{equation}
where the second term encodes quantum corrections. Substituting \eqref{eq:rho_APS} into \eqref{eq:main_mapping} yields the quantum-corrected Schwarzschild metric~\cite{Kelly:2020lec, Ali:2022lec, Lewandowski:2023lec}
\begin{equation}
	f_{\text{q-Sch}}(R) = 1 - \frac{2M}{R} + \frac{4\ell^2 M^2}{R^4}, \qquad \ell^2 = \frac{3}{8\pi\rho_c},
	\label{eq:f_qSch}
\end{equation}
with $\ell^2 = \gamma^2\Delta$ setting the short-distance scale ($\gamma$: Barbero--Immirzi parameter; $\Delta$: area gap). Charged and cosmological extensions follow immediately via the universal substitution $\rho_{\text{m}} \to \rho_{\text{s}}$ in \eqref{eq:rho_APS}:
\begin{equation}
	f_{\text{q-RN}}(R) = f_{\text{RN}}(R) + \frac{\ell^2}{R^2}\bigl[1 - f_{\text{RN}}(R)\bigr]^2.
	\label{eq:f_qRN}
\end{equation}
This algebraic extension reconstructs the geometry directly from the density ansatz, avoiding the need to solve the full set of modified differential field equations.

\subsection{From geometry to density.} Starting from a known $f(R)$ reveals the physical content of regular black holes. Quantum-corrected geometries can also arise from classical NLED, which introduces an extra degree of freedom. The LQG-corrected Schwarzschild metric \eqref{eq:f_qSch} is reproduced by the NLED Lagrangian~\cite{AyonBeato:1998}
\begin{equation}
	\mathcal{L}(F) = -\frac{\eta}{16\pi}F^{3/2}, \qquad \eta > 0,
\end{equation}
with purely magnetic field $F_{\theta\phi}=Q_m\sin\theta$. The energy density is $\rho(R) = \eta\sqrt{2}\,Q_m^3/(8\pi R^6)$. Integrating $m'(R)=4\pi R^2\rho(R)$ gives
\begin{equation}
	f(R) = 1 - \frac{2M}{R} + \frac{\eta\sqrt{2}\,Q_m^3}{3R^4},
	\label{eq:f_NLED}
\end{equation}
Matching to~\eqref{eq:f_qSch} requires $\eta\sqrt{2}\,Q_m^3/3 = 4\ell^2 M^2$, or $Q_m=(12\ell^2/(\eta\sqrt{2}))^{1/3}M^{2/3}$, relating the magnetic charge $Q_m$ to the quantum scale $\ell^2$ (see Appendix A). Thus identical geometries can emerge from distinct physical origins.

The Hayward metric~\cite{Hayward:2005gi}
\begin{equation}
	f_{\text{Hay}}(R) = 1 - \frac{2M/R}{1 + 2M\tilde{\ell}^2/R^3}
	\label{eq:f_Hayward}
\end{equation}
maps to the effective density
\begin{equation}
	\rho_{\text{Hay}}(R) = \rho_{\text{m}}\left(1 + \frac{\rho_{\text{m}}}{\tilde{\rho}_c}\right)^{-1}, \qquad \tilde{\ell}^2 = \frac{3}{8\pi\tilde{\rho}_c}.
	\label{eq:rho_Hayward}
\end{equation}
In the limit $\tilde{\ell}^2 \ll 1$, this agrees with the APS form \eqref{eq:rho_APS} at leading order, showing that Hayward gravity and LQG corrections share the same leading-order behavior. Applying $\rho_{\text{m}} \to \rho_{\text{s}}$ to \eqref{eq:rho_Hayward} yields the charged Hayward--(A)dS solution
\begin{equation}
	f_{\text{Hay-ch}}(R) = 1 - \bigl[1 - f_{\text{RN}}(R)\bigr]\left(1 + \frac{\tilde{\ell}^2}{R^2}\bigl[1 - f_{\text{RN}}(R)\bigr]\right)^{-1},
	\label{eq:f_Hayward_charged}
\end{equation}
which agrees at leading order with \eqref{eq:f_qRN} when $\tilde{\ell}^2 \ll 1$.

The Bardeen metric~\cite{Bardeen1968}
\begin{equation}
	f_{\text{Bar}}(R) = 1 - \frac{2M}{R}\left(1 + \frac{g^2}{R^2}\right)^{-3/2}
	\label{eq:f_Bardeen}
\end{equation}
corresponds to
\begin{equation}
	\rho_{\text{Bar}}(R) = \rho_{\text{m}}\left[1 + \left(\frac{\rho_{\text{m}}}{\bar{\rho}_c}\right)^{2/3}\right]^{-3/2}, \qquad g^2 = \left(\frac{3M}{4\pi\bar{\rho}_c}\right)^{2/3}.
	\label{eq:rho_Bardeen}
\end{equation}
Expanding for small $g^2 \ll 1$ gives
\begin{equation}
	\rho_{\text{Bar}}(R) = \rho_{\text{m}}\left(1 - \frac{3}{2}\left(\frac{\rho_{\text{m}}}{\bar{\rho}_c}\right)^{2/3} + \cdots\right),
	\label{eq:rho_Bardeen_expansion}
\end{equation}
which involves fractional powers of $\rho_{\text{m}}$. Unlike the APS and Hayward densities---which admit Taylor expansions in integer powers of $\rho_{\text{m}}/\rho_c$---the Bardeen density contains fractional, noninteger powers. This indicates that the Bardeen solution originates from a phenomenological model. Nevertheless, the mapping applies universally: substituting $\rho_{\text{m}} \to \rho_{\text{s}}$ in \eqref{eq:rho_Bardeen} yields the charged Bardeen--(A)dS solution without solving field equations.

In summary, the mapping \eqref{eq:main_mapping} provides a unified algebraic framework for generating black hole solutions and predicting their collapse dynamics across classical and quantum theories.

\section{Integer-order corrections as a diagnostic of systematic UV-motivated expansions} 

The structure of density corrections can diagnose whether a metric is compatible with a systematic UV-motivated expansion or is more naturally interpreted phenomenologically. Writing the total effective density as
\begin{equation}
	\rho(R) = \rho_{\text{m}} \, \mathcal{F}\!\left(\frac{\rho_{\text{m}}}{\rho_c}\right),
	\label{eq:rho_general}
\end{equation}
a correction is of \textit{integer order} precisely when $\mathcal{F}$ admits a Taylor expansion in integer powers of the dimensionless ratio $x = \rho_{\text{m}}/\rho_c$:
\begin{equation}
	\mathcal{F}(x) = 1 + c_1 x + c_2 x^2 + c_3 x^3 + \cdots.
	\label{eq:F_expansion}
\end{equation}
Within the ansatz \eqref{eq:rho_general}, this pattern is equivalent to requiring that the metric function contains exclusively terms of the form $R^{-(3n-2)}$ in $1-f(R)$:
\begin{equation}
	1 - f(R) = \frac{2M}{R} + \frac{a_2 M^2}{R^4} + \frac{a_3 M^3}{R^7} + \frac{a_4 M^4}{R^{10}} + \cdots.
	\label{eq:integer_structure}
\end{equation}
We emphasize that this diagnostic identifies theories with a \textit{systematic perturbative expansion} in a dimensionless coupling; non-analytic terms may arise from non-perturbative effects, and integer-order expansions can appear in effective theories. The criterion is most powerful when combined with independent theoretical constraints. Each power $R^{-(3n-2)}$ maps directly to a density correction proportional to $\rho_{\text{m}}^{\,n}$, reflecting a systematic expansion in a dimensionless coupling. This integer-order pattern is preserved under the universal substitution $\rho_{\text{m}} \to \rho_{\text{s}}$, where $\rho_{\text{s}} = 3M/(4\pi R^3) - 3q^2/(8\pi R^4) + \Lambda/(8\pi)$ incorporates dust, charge, and cosmological constant contributions. The diagnostic thus applies to charged and asymptotically (A)dS extensions as an expansion in $\rho_{\text{s}}$; the simple inverse-radius test $R^{-(3n-2)}$ should be understood as the neutral limit.

LQC provides the prototype: the Ashtekar--Pawlowski--Singh density yields the quantum-corrected Schwarzschild metric with only the $R^{-4}$ term at leading order. Higher holonomy corrections generate $R^{-7}$, $R^{-10}$, and so on, preserving the integer structure throughout. Applying $\rho_{\text{m}} \to \rho_{\text{s}}$ extends this directly to the quantum-corrected RN--(A)dS solution~\eqref{eq:f_qRN}, which retains the same pattern. Similarly, generalized quasi-topological gravity (e.g., Einsteinian cubic gravity~\cite{Bueno:2016xff}) and regularized Einstein--Gauss--Bonnet gravity~\cite{Glavan:2019inb} produce metrics of the form~\eqref{eq:integer_structure}; their charged/(A)dS extensions inherit the same structure. The Hayward metric also follows this pattern: its expansion contains only $R^{-(3n-2)}$ terms, and for small $\tilde{\ell}^2$ it agrees at leading order with the LQG correction.

In contrast, the Bardeen black hole introduces $R^{-3}$ and $R^{-5}$ terms in $1-f(R)$, violating~\eqref{eq:integer_structure}. This yields fractional powers in the density expansion---specifically $(\rho_{\text{m}}/\bar{\rho}_c)^{2/3}$---signaling a phenomenological origin. Even after extending to charged solutions via $\rho_{\text{m}} \to \rho_{\text{s}}$, the fractional powers persist in charged/(A)dS extensions, supporting its interpretation as an effective phenomenological model rather than as a simple perturbative UV expansion. Within this diagnostic, any metric failing to satisfy this integer-order structure is best regarded as an effective or phenomenological model rather than as the low-energy expansion of a systematic UV-motivated expansion.

This diagnostic is independent of the collapse outcome. Both integer-order and fractional corrections can produce bouncing or soft-landing collapse, depending solely on $f(R)$. Our mapping therefore provides a simple criterion: given any static solution $f(R)$, expand $1-f(R)$ in inverse powers of $R$. If every term follows $R^{-(3n-2)}$, the solution is compatible with a systematic UV-motivated expansion; otherwise, it is more naturally interpreted as phenomenological. Conversely, the presence of fractional powers does not automatically invalidate a model; it suggests that the ansatz may be phenomenological or that the expansion parameter is not simply $\rho_{\rm m}/\rho_c$. Such models can still provide valuable effective descriptions of high-curvature physics.

\section{Geometric classification of collapse dynamics}

The collapse dynamics are entirely governed by the exterior metric function $f(R)$ in Eq.~\eqref{eq:surface_law} through $1 - f(R) \geq 0$. This leads to three possible outcomes~\cite{Li:2026OS}: \textit{singular} collapse, where the surface reaches $R=0$ in finite proper time; \textit{bouncing} collapse, where $f(R_*)=1$ and $f'(R_*)<0$ at finite $R_*>0$; and \textit{soft-landing} collapse, where $f(R)<1$ for all $R>0$ but $f(R)\to 1$ as $R\to 0$ (with true curvature regularity requiring a de~Sitter-like core~\cite{Li:2023yyw}). In the bouncing case, consistency of the trapped region requires $R_*<R_-$ and, more strictly, the apparent-horizon minimum to lie inside the inner horizon.

Moreover, physical consistency imposes strict hierarchies between the apparent-horizon minimum and the event horizons. For example, in RN-like collapse, requiring the minimum to lie inside the inner horizon yields parameter bounds analogous to Eq.~(30) in Ref.~\cite{Li:2026OS}. The collapse class is determined not by the functional \textit{form} of $f(R)$, but by its \textit{parameter values}. The Hayward metric~\eqref{eq:f_Hayward}, for example, realizes bouncing behavior for $\tilde\ell^2\ll M^2$ but transitions to soft-landing as $\tilde\ell^2$ increases (see Appendix C). More generally, distinct physical origins can produce identical collapse dynamics when $f(R)$ coincides. The LQG-corrected metric admits an exact classical counterpart in Einstein gravity coupled to NLED~\cite{AyonBeato:1998} under the parameter matching $\eta\sqrt{2}\,Q_m^3/3 = 4\ell^2 M^2$. When satisfied, quantum and classical descriptions yield indistinguishable surface trajectories, apparent-horizon evolution, and horizon structure. 

Both descriptions respect cosmic censorship. After bouncing at $R_* < R_-$, the surface moves outward toward the inner horizon $R_-$. Generic perturbations near $R_-$ trigger mass inflation~\cite{Poisson:1990eh}, producing a weak null singularity that replaces the Cauchy horizon. The classical $R=0$ singularity remains causally disconnected, preserving weak cosmic censorship~\cite{Penrose:1969pc}. The mass-inflation singularity at $R_-$ upholds strong cosmic censorship~\cite{Sorce:2017dst}. Thus, the physically relevant endpoint occurs near $R_-$, not at the center (see Appendix C).

\section{Conclusions and outlook}

We have established an exact bidirectional mapping between the effective homogeneous interior density of a collapsing star and the exterior metric function within the force-free generalized OS framework. Inserting an effective density profile into the Friedmann equation reconstructs the corresponding candidate static geometry and surface dynamics. The result should not be interpreted as a universal reconstruction of arbitrary quantum-corrected collapse, but as an exact algebraic method for the subset of models admitting a homogeneous interior, no thin shell, and a static exterior. This approach works uniformly across LQG, quasi-topological gravity, nonlinear electrodynamics, and regular black hole models. This does not imply that all quantum-corrected or modified-gravity collapse scenarios admit a homogeneous interior or a static exterior; rather, it identifies the subset for which such an exact matching description is available.

Two points emerge clearly. First, the fate of collapse---singular, bouncing, or soft-landing---is fixed by the exterior metric function $f(R)$. This is illustrated by the equivalence between LQG corrections and classical nonlinear electrodynamics: with matched parameters, the resulting surface trajectories and horizon evolution are the same. Second, the mathematical structure of density corrections distinguishes systematic UV-motivated perturbative expansions from more phenomenological models. Integer-order expansions in $\rho_{\text{m}}/\rho_c$---producing metric terms $R^{-(3n-2)}$---are characteristic of the former, while fractional powers, as in the Bardeen black hole, indicate the latter. Within the GOS matching framework, Eq.~\eqref{eq:main_mapping} reduces the differential matter–geometry relation of Einstein's field equations to a much simpler purely algebraic correspondence.

Finally, we remark that our algebraic mapping complements Hamiltonian-based reconstruction methods~\cite{ZhangCao2025}. There, covariant vacuum theories are reconstructed from a prescribed family of static, spherically symmetric metrics; here, within the GOS matching framework, the exterior geometry and collapse dynamics are reconstructed algebraically from interior effective density corrections, and conversely the effective density is extracted from any chosen exterior metric. The two approaches thus address distinct layers of the same broader question: the covariant dynamics underlying a vacuum geometry versus the matter content driving a dynamical collapse.

Future extensions to time-dependent exteriors, inhomogeneous interiors, and shell-supported junctions would be necessary to address the most general quantum-corrected collapse problem. The present article isolates the exactly solvable force-free sector, providing a rigorous baseline. Extensions of our approach to rotating collapse and inhomogeneous interiors, where thin shells or anisotropic stresses may emerge, remain important directions for future work.
	
\section*{Acknowledgement}

This work is supported in part by the National Natural Science Foundation of China (NSFC) Grant Nos.~W2533015, 12375052, and 11935009 and by the Tianjin University Self-Innovation Fund for Extreme Basic Research Grant No.~2025XJ21-0007.

\section*{Appendix}

\begin{appendix}

\section{Classical realization of quantum-corrected geometry via nonlinear electrodynamics}
\label{app:NLED}

We emphasize that NLED serves here as a classical completion engineered to reproduce a prescribed $f(R)$, with the magnetic charge providing an additional degree of freedom. It is not intended as a fundamental electromagnetic theory with a Maxwell weak-field limit. Consider the action
\begin{equation}
	S = \int \dd^4x\,\sqrt{-g}\left[\frac{\mathcal{R}}{16\pi} + \mathcal{L}(F)\right],
\end{equation}
where $F = F_{\mu\nu}F^{\mu\nu}$ and $\mathcal{R}$ is the Ricci scalar. The stress tensor and generalized Maxwell equation read
\begin{equation}
	T_{\mu\nu} = -4\mathcal{L}_F\,F_{\mu\alpha}F_{\nu}{}^{\alpha}+g_{\mu\nu}\mathcal{L},
	\qquad
	\nabla_\mu(\mathcal{L}_F F^{\mu\nu}) = 0,
	\label{eq:NLED_Tmunu}
\end{equation}
where $\mathcal{L}_F \equiv \dd\mathcal{L}/\dd F$. We adopt a purely magnetic, spherically symmetric configuration $F_{\theta\phi}=Q_m\sin\theta$, which respects spherical symmetry and describes a conserved magnetic charge. For this configuration the invariant is $F=2Q_m^2/R^4>0$. Purely magnetic fields satisfy $T^t{}_t = T^R{}_R = \mathcal{L}$ and $T^\theta{}_\theta = T^\phi{}_\phi = \mathcal{L} - 2F\mathcal{L}_F$, giving
\begin{equation}
	\rho = -\mathcal{L},
	\qquad
	p_r = -\rho,
	\qquad
	p_t = \mathcal{L} - 2F\mathcal{L}_F .
	\label{eq:NLED_rho_pr_pt_general}
\end{equation}

Since $F \propto R^{-4}$ for a magnetic monopole, choosing $\mathcal{L} \propto -F^{3/2}$ yields
\begin{equation}
	\mathcal{L}(F) = -\frac{\eta}{16\pi}\,F^{3/2}, \qquad \eta > 0,
	\label{eq:NLED_L_F32}
\end{equation}
giving the energy density and pressures
\begin{equation}
	\rho(R) = -p_r(R) = \frac{p_t(R)}{2} = \frac{\eta\sqrt{2}\,Q_m^3}{8\pi R^6}, 
	\label{eq:NLED_rho_pr_pt}
\end{equation}
where we take $Q_m>0$ without loss of generality. The weak and strong energy conditions are satisfied everywhere, while the dominant energy condition is violated ($p_t=2\rho>\rho$). However, because $\rho\propto R^{-6}$ decays rapidly, this violation is confined to the high-curvature core and has negligible macroscopic impact. Such localized dominant-energy-condition violation is a known feature of many regular black-hole models sourced by nonlinear electrodynamics~\cite{AyonBeato:1998,Bronnikov:2001}. Since the violation is confined to the high-curvature core ($R \ll M$) and the weak/strong energy conditions hold everywhere, the macroscopic dynamics and junction conditions remain physically consistent. Moreover, because the bounce occurs at $R_* < R_-$ while remaining outside the high-curvature core where the surface-supported electromagnetic source dominates, the effective description at the matching surface $\Sigma$ is insensitive to the detailed high-curvature behavior.

Writing the metric function as $f(R)=1-2m(R)/R$, Einstein's equations yield the Misner--Sharp mass relation $m'(R)=4\pi R^2\rho(R)$~\cite{Misner:1964}. Substituting Eq.~\eqref{eq:NLED_rho_pr_pt} gives $m'(R)=\eta\sqrt{2}\,Q_m^3/(2R^4)$, which integrates to
\begin{equation}
	m(R)=M-\frac{\eta\sqrt{2}\,Q_m^3}{6R^3},
	\qquad
	f(R)=1-\frac{2M}{R}+\frac{\eta\sqrt{2}\,Q_m^3}{3R^4}.
	\label{eq:NLED_f_of_R}
\end{equation}
Matching to the quantum-corrected form \eqref{eq:f_qSch} requires
\begin{equation}
	\frac{\eta\sqrt{2}\,Q_m^3}{3}=4\ell^2 M^2,
	\qquad\text{or}\qquad
	Q_m=\left(\frac{12\,\ell^2}{\eta\sqrt{2}}\right)^{1/3}M^{2/3}.
	\label{eq:Qm_in_terms_of_alpha}
\end{equation}
Thus the magnetic charge scales as $Q_m\propto M^{2/3}$, requiring mass-dependent tuning to reproduce the same exterior geometry.

The identical spacetime therefore admits two distinct interpretations. In the quantum picture, short-distance repulsion comes from LQG corrections encoded in $\ell^2$. In the classical picture, the same geometry emerges as an exact solution of Einstein gravity coupled to NLED, where a magnetic monopole provides the repulsive core. The additional degree of freedom $Q_m$ can be tuned to reproduce precisely the $R^{-4}$ correction of the quantum-corrected metric, yielding identical exterior geometry without including quantum effects.

\section{Quantum-corrected Oppenheimer--Snyder collapse dynamics}
\label{app:qOS}

We model gravitational collapse as a homogeneous dust interior ($k=0$ FRW) matched to a static spherically symmetric exterior under force-free Darmois--Israel junction conditions. The exterior metric reads
\begin{equation}
	ds^2_{\rm out} = -f(R)\,dt^2 + \frac{dR^2}{f(R)} + R^2 d\Omega^2,
	\label{eq:static_out}
\end{equation}
and the star's surface at fixed comoving radius $r_\Sigma$ has areal radius $R(\tau)=a(\tau)r_\Sigma$, where $\tau$ is the proper time of comoving interior observers. Working in Painlevé--Gullstrand coordinates, which are regular across horizons and synchronized with $\tau$, the junction conditions yield the surface evolution law
\begin{equation}
	\dot{R}^2 = 1 - f(R),
	\label{eq:surface_law_app}
\end{equation}
where the collapsing branch takes $\dot{R}<0$. Real motion requires $1-f(R)\geq0$. In the absence of a thin shell, the junction conditions require continuity of radial stress across $\Sigma$, which is automatically satisfied by the mapping \eqref{eq:main_mapping}.

The surface trajectory follows from integrating Eq.~\eqref{eq:surface_law_app}:
\begin{equation}
	T(R) = \int_R^{R_0} \frac{dR'}{\sqrt{1-f(R')}},
	\label{eq:T_of_R}
\end{equation}
with $R_0$ the initial radius. The apparent horizon inside the star is
\begin{equation}
	R_{\rm AH}(R) = \frac{R}{\sqrt{1-f(R)}},
	\label{eq:RAH}
\end{equation}
which diverges at a bounce where $f(R_\ast)=1$. The interior event horizon, obtained by tracing outgoing null geodesics backward from the outer horizon $R_+$ (where $f(R_+)=0$), is given by
\begin{equation}
	R_{\rm eh}(R) = R\left(1 - \int_{R_+}^{R} \frac{dR'}{R'\sqrt{1-f(R')}}\right),
	\label{eq:Reh}
\end{equation}
satisfying $R_{\rm eh}(R_+)=R_+$ and $R_{\rm eh}(R)<R$ for $R>R_+$.

Equations \eqref{eq:surface_law_app}--\eqref{eq:Reh} demonstrate that the entire collapse dynamics and horizon structure depend only on the exterior metric function $f(R)$. This permits two equivalent approaches: (i) begin with an effective interior dynamics (e.g., LQC) and reconstruct $f(R)$, or (ii) specify a candidate exterior $f(R)$ and derive the induced surface evolution. When the Darmois--Israel conditions hold without a thin shell, both descriptions are dynamically equivalent on the surface $\Sigma$.

We now analyze the collapse dynamics for the quantum-corrected metric
\begin{equation}
	f_{\rm qOS}(R) = 1 - \frac{2M}{R} + \frac{\alpha M^2}{R^4}, \qquad \alpha \equiv 4\ell^2,
	\label{eq:f_qOS}
\end{equation}
which arises both from LQG corrections~\cite{Kelly:2020lec, Ali:2022lec, Lewandowski:2023lec} and from the NLED realization (Appendix A) when $\alpha = \eta\sqrt{2}\,Q_m^3/(3M^2)$. This example concretely illustrates bouncing collapse without repeating analyses of RN or regular black holes from prior work~\cite{Li:2026OS}.

Under force-free junction conditions, the star's surface evolves via $\dot{R}^2 = 1 - f_{\rm qOS}(R)$ with $\dot{R} < 0$ during collapse. Integrating yields the proper time to radius $R$:
\begin{equation}
	T(R) = \frac{1}{3\sqrt{M}}\left[\sqrt{2R_0^3 - \alpha M} - \sqrt{2R^3 - \alpha M}\right],
	\label{eq:T_qOS}
\end{equation}
where $R_0$ is the initial radius. The bounce occurs at $R_* = (\alpha M/2)^{1/3}$ where $\dot{R}=0$ and $\ddot{R}>0$, with bounce time $T_* = T(R_*)$.

Three characteristic radii govern the dynamics, forming the strict hierarchy $R_{\rm infl} > R_{\rm turn} > R_*$:
\begin{align}
	R_*^3 &= \frac{\alpha M}{2}, & (\dot{R}=0,\ \text{bounce}), \nonumber\\
	R_{\rm turn}^3 &= \alpha M = 2R_*^3, & (dR_{\rm AH}/dT=0,\ \text{apparent-horizon vertex}), \nonumber\\
	R_{\rm infl}^3 &= 2\alpha M = 4R_*^3, & (\ddot{R}=0,\ \text{inflection point}).
\end{align}
The surface accelerates for $R > R_{\rm infl}$ and decelerates for $R_* < R < R_{\rm infl}$. The apparent horizon radius is $R_{\rm AH}(R) = R / \sqrt{1 - f_{\rm qOS}(R)}$, which diverges at $R_*$.

Horizons satisfy $f_{\rm qOS}(R) = 0$, i.e., $R^4 - 2MR^3 + \alpha M^2 = 0$. A degenerate case ($f = f' = 0$) occurs at $\alpha_{\rm crit} = 27M^2/16$ with $R^{(\rm deg)} = 3M/2$. For $0 < \alpha < \alpha_{\rm crit}$, two horizons exist; for $\alpha > \alpha_{\rm crit}$, none. In the weak-correction regime $\alpha / M^2 \ll 1$,
\begin{equation}
	R_- \simeq R_*, \qquad R_+ \simeq 2M - \frac{\alpha}{8M},
\end{equation}
with $R_- > R_*$ holding exactly.

Physical consistency of the trapped region requires the apparent-horizon minimum to form inside the inner horizon: $R_{\rm turn} < R_-$. Using $f_{\rm qOS}(R_{\rm turn}) = 1 - M/R_{\rm turn}$, the boundary case $R_{\rm turn} = R_-$ gives $R_{\rm turn} = M$ and $\alpha = M^2$. Thus, consistent apparent-horizon minimum formation requires
\begin{equation}
	\alpha < M^2 \quad (\text{in addition to } \alpha < \alpha_{\rm crit}),
\end{equation}
which translates via the NLED matching condition \eqref{eq:Qm_in_terms_of_alpha} into an upper bound on the magnetic charge $Q_m$.

\begin{figure}[t]
	\centering
	\includegraphics[width=0.5\linewidth]{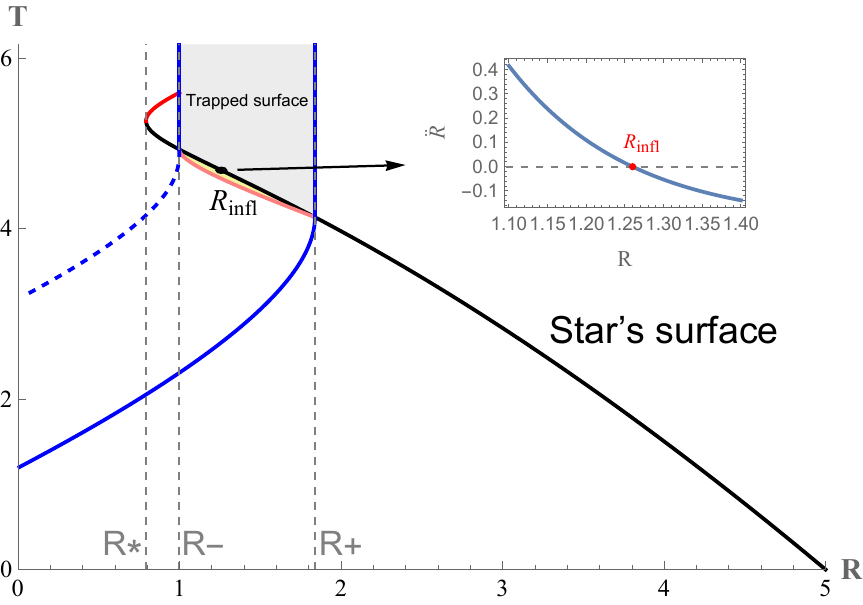}
	\caption{Surface trajectory $R(T)$ for gravitational collapse with $M=1$ (in Planck units), initial radius $R_0=5$, and quantum correction parameter $\alpha=1$. The star bounces at $R_* \simeq 0.79$ between the inner and outer horizons $R_- \simeq 1.00$ and $R_+ \simeq 1.84$. Inset: proper acceleration $\ddot{R}$ changes sign at the inflection point $R_{\rm infl} \simeq 1.26$, marking the transition from acceleration to deceleration.}
	\label{fig:dynamics}
\end{figure}

Following the bounce, the surface evolves toward $R_-$ where mass inflation~\cite{Poisson:1990eh} generically produces a weak null singularity. The classical $R=0$ singularity remains causally disconnected, preserving weak cosmic censorship~\cite{Penrose:1969pc}. The mass-inflation singularity at $R_-$ upholds strong cosmic censorship~\cite{Sorce:2017dst} by converting the Cauchy horizon into a curvature singularity. Hence, the physically relevant endpoint occurs near $R_-$.

This analysis confirms that the entire collapse dynamics---surface trajectory \eqref{eq:T_of_R}, apparent horizon \eqref{eq:RAH}, event horizon \eqref{eq:Reh}, characteristic radii, and cosmic censorship implications---depends solely on $f_{\rm qOS}(R)$. Whether derived from quantum geometry or classical NLED, identical $f(R)$ yields identical physics. The mapping framework thus provides a geometric lens: singularity resolution is encoded in spacetime structure alone, independent of microscopic origin.

\section{Geometric classification and dynamical consistency}
\label{app:classification}

This section provides the rigorous derivations supporting the collapse classification discussed in the main text.

\paragraph{Hierarchy of characteristic radii.} For bouncing collapse, the turning point $R_*$ satisfies $f(R_*)=1$ and $f'(R_*)<0$. Physical motion requires $1-f(R)\geq 0$, so the region $0<R<R_*$ is classically forbidden and implies $f(R)>1$. Suppose, for contradiction, that $R_* \geq R_-$, where $R_-$ is the inner horizon satisfying $f(R_-)=0$. Since $R_-$ would then lie in the forbidden region, we would have $f(R_-)>1$, directly contradicting the horizon definition. Hence $R_* < R_-$ strictly. This hierarchy holds for any static metric with two horizons. In the RN-type interpretation used here, the FLRW bulk remains neutral and the charge is carried by the surface $\Sigma$; hence there is no charged FLRW core. The neutral-dust assumption refers to the homogeneous bulk interior and is not spoiled by the surface charge that sources the exterior RN field. For broader horizon-counting constraints under energy conditions, see Ref.~\cite{Yang:2021horizons}.

Physical consistency of the trapped region further requires the apparent-horizon minimum to form inside the inner horizon: $R_{\rm turn} < R_-$. Using the quantum-corrected metric~\eqref{eq:f_qOS}, we have $f_{\rm qOS}(R_{\rm turn}) = 1 - M/R_{\rm turn}$. Imposing the boundary case $R_{\rm turn} = R_-$ yields $R_{\rm turn} = M$ and $\alpha = M^2$. Thus, consistent apparent-horizon formation requires
\begin{equation}
	\alpha < M^2 \quad (\text{in addition to } \alpha < \alpha_{\rm crit}),
	\label{eq:alpha_bound_SM}
\end{equation}
which translates via the NLED matching condition~\eqref{eq:Qm_in_terms_of_alpha} into an upper bound on the magnetic charge $Q_m$. Numerical verification of this hierarchy is shown in Fig.~\ref{fig:dynamics}.

\paragraph{Hayward soft-landing and asymptotic expansion.} Expanding the Hayward metric~\eqref{eq:f_Hayward} yields
\begin{equation}
	1 - f_{\text{Hay}}(R) = \frac{2M}{R} - \frac{4M^2\tilde{\ell}^2}{R^4} + \frac{8M^3\tilde{\ell}^4}{R^7} - \cdots.
	\label{eq:Hayward_exp_SM}
\end{equation}
This expansion reproduces the LQG-corrected form~\eqref{eq:f_qSch} only at leading order in the large-$R$ expansion. It should not be interpreted as a finite-radius bounce of the full Hayward geometry. Indeed, for the full rational metric, $1-f_{\text{Hay}}(R)=2M/[R(1+2M\tilde{\ell}^2/R^3)]>0$ for all $R>0$, so $f_{\text{Hay}}(R)=1$ has no positive finite-radius root. The Hayward metric therefore realizes soft-landing collapse, with $f_{\text{Hay}}(R)<1$ for all $R>0$ and $f_{\text{Hay}}(R)\to1$ as $R\to0$. Note that $f(0)=1$ alone does not ensure curvature regularity; some higher-derivative models (e.g., Einstein--Gauss--Bonnet~\cite{Li:2026OS}) satisfy this condition yet retain a central singularity. True regularity demands a de~Sitter-like core near $R=0$ with $\rho+p_r=0$ and $\rho+p_t>0$, satisfying the weak and strong energy conditions~\cite{Li:2023yyw}.

\paragraph{Mass inflation and cosmic censorship.} After bouncing at $R_* < R_-$, the surface evolves toward the inner horizon. Generic perturbations near $R_-$ trigger mass inflation~\cite{Poisson:1990eh}, producing a weak null singularity that replaces the Cauchy horizon. The classical $R=0$ singularity remains causally disconnected, preserving weak cosmic censorship~\cite{Penrose:1969pc}. The mass-inflation singularity at $R_-$ upholds strong cosmic censorship~\cite{Sorce:2017dst} by converting the Cauchy horizon into a curvature singularity. Hence the physically relevant dynamical endpoint occurs near $R_-$, not at the center. This mechanism applies universally to any $f(R)$ yielding bouncing dynamics.

\end{appendix}
	
\bibliographystyle{JHEP}

\end{document}